\documentclass[11pt]{article}

\usepackage{amssymb,amsmath,amsfonts,amsthm,lscape}

\usepackage{hyperref}
\hypersetup{
   colorlinks   =  true,
    linkcolor    = cyan,
    citecolor    = red,
     urlcolor	=magenta,     
}

\newcommand\be{\begin{equation}}
\newcommand\ee{\end{equation}}
\newcommand\bea{\begin{eqnarray}}
\newcommand\eea{\end{eqnarray}}

\newcommand\kk{{\eta}} 
  
\DeclareMathOperator{\spn}{span}

\begin{document}

\thispagestyle{empty}

\

 \vskip1cm

\noindent {\Large{\bf {Drinfel'd double structures for Poincar\'e and Euclidean groups
}}}

\medskip
\medskip
\medskip
\medskip

\begin{center}

{\sc Ivan Gutierrez-Sagredo\footnote{
 Based on the contribution presented at ``The 32nd International Colloquium on Group Theoretical Methods in Physics" (Group32),
July 9--13, 2018, 
 Prague, Czech Republic.}, Angel Ballesteros and Francisco  J  Herranz}

\medskip
\medskip

{Departamento de F\1sica, Universidad de Burgos, E-09001 Burgos, Spain}

 {E-mails:\quad {\tt  igsagredo@ubu.es, angelb@ubu.es,  fjherranz@ubu.es}}

 \end{center}

\begin{abstract}  
\noindent
All non-isomorphic three-dimensional Poisson homogeneous Euclidean spaces are constructed and analyzed, based on the classification of coboundary Lie bialgebra structures of the Euclidean group in 3-dimensions, and the only Drinfel'd double structure for this group is explicitly given. The similar construction for the Poincar\'e case is reviewed and the striking differences between the Lorentzian and Euclidean cases are underlined. Finally, the contraction scheme starting from Drinfel'd double structures of the $\mathfrak{so}(3,1)$ Lie algebra is presented.
\end{abstract}

\medskip\medskip

\section{Introduction}

The search for a quantum theory of gravity is nowadays one of the most challenging problems in theoretical physics. With this motivation various either partial or schematic models have been proposed from different perspectives, with the hope of getting useful insights in certain important aspects of the full theory, in particular those related to the interplay between quantum and gravitational properties of spacetime at the Planck scale. One such a particularly interesting approach is that of (2+1)-dimensional quantum gravity \cite{Carlip2003book}. In this case the main feature of the theory is provided by the fact that general relativity in (2+1)-dimensions is a topological  theory, and solutions of the vacuum Einstein field equations are locally isometric to one of the three maximally symmetric spacetimes of constant curvature, thus allowing a description that strongly depends on the sign of the cosmological constant. This fact greatly simplifies the picture and allows for several quantization approaches. In some of them, quantum groups play a relevant role, like in state sum models or spin foams (see for instance~\cite{TV1992invariants,AGS1995,BR1995,MS2003} and references therein). 

The description of (2+1)-gravity as a Chern-Simons theory, in which the group of isometries of the corresponding spacetime model plays the role of the gauge group \cite{AT1986,Witten1988}, will be specially relevant in what follows. As it was proved in \cite{AM1995moduli,FR1999moduli}, the phase space structure of this theory is related with the moduli space of flat connections on a Riemann surface. Symmetries on this phase space are described by Poisson-Lie groups, and this means that quantum groups turn out to be the natural  candidates for symmetries once the quantization of the theory had been performed. In \cite{FR1999moduli} it was proved that the relevant Poisson-Lie structures are those defined by classical $r$-matrices fulfilling the so-called Fock-Rosly condition, which was proven in \cite{BHM2013cqg} to be automatically verified if the chosen $r$-matrix is the canonical one coming from a Drinfel'd double (DD) structure of the corresponding model Lie group of isometries. Therefore, the study of DD structures and their quantization (see~\cite{MW1998, BM1998topological, BM2003, Majid2005time, JMN2009}) turns out to be a relevant subject in a quantum gravity context, and recent works have been devoted to the explicit construction and analysis of DD structures for the Poincar\'e group \cite{BGH2018poincareDD} and for the  (anti-)de Sitter ones \cite{BHM2013cqg}. Once the DDs are classified, the natural question that arises concerns which are the noncommutative spacetimes obtained as quantizations of the corresponding Poisson homogeneous spaces induced by the DDs, and which are their main structural features (see, for instance, \cite{BHM2010plb, BHM2014plb, BMN2017homogeneous, BGHMN2017dice} and references therein). This question has been fully answered for the Poincar\'e case in \cite{BGH2018poincareDD}, and in this work we provide the Euclidean counterparts of these results, given that spaces with Euclidean signature are also commonly employed in quantum gravity considerations, as shown in \cite{MS2009semidualization}. These results will be put into correspondence with the full classification of Euclidean $r$-matrices given in~\cite{Stachura1998} (note that analogous classifications for the (anti-)de Sitter cases can be found in~\cite{BorowiecLukierskiTolstoy2016rmatrices,BorowiecLukierskiTolstoy2016rmatricesaddendum}), and the analysis of the contraction to the Euclidean case from the DDs for the Lie algebra $\mathfrak{so}(3,1)$, which were provided in~\cite{BHM2013cqg}, will be also given. 


\section{Drinfel'd doubles, Lie bialgebras and Poisson homogeneous spaces}

In this section we quickly review the mathematical structures needed for the rest of the work. For further details see \cite{BGH2018poincareDD,BMN2017homogeneous} and references therein. The first structure we need is that of a {\em(classical) Drinfel'd double}:  A $2d$-dimensional Lie algebra $\mathfrak{a}$ is said to be the Lie algebra of  a (classical) DD Lie group~\cite{Drinfeld1987icm} if there exists a basis $\{Y_1,\dots,Y_d,y^1,\dots,y^d \}$ of $\mathfrak a$ in which the 
Lie bracket reads
\begin{align}
[Y_i,Y_j]= c^k_{ij}Y_k, \qquad  
[y^i,y^j]= f^{ij}_k y^k, \qquad
[y^i,Y_j]= c^i_{jk}y^k- f^{ik}_j Y_k.  \, \label{agd}
\end{align}
We call $\mathfrak{g}=\mbox{span}\{Y_1,\dots,Y_d\}$ and $\mathfrak{g}^\ast=\mbox{span}\{y^1,\dots,y^d \}$. The (classical) DD is the unique connected and simply connected Lie group $G$ with Lie algebra given by $\mathfrak{a}$. The Lie algebra $\mathfrak{a}$ is said to be the double Lie algebra of $\mathfrak{g}$ and of its dual Lie algebra $\mathfrak{g}^\ast$, where the duality 
is defined with respect to the nondegenerate symmetric bilinear form $\langle \, , \, \rangle : \mathfrak{a} \times \mathfrak{a} \rightarrow \mathbb{R}$ given by
\begin{align}\label{ages}
 \langle Y_i,Y_j\rangle= 0,\qquad \langle y^i,y^j\rangle=0, \qquad
\langle y^i,Y_j\rangle= \delta^i_j,\qquad \forall i,j  ,
\end{align}
which is `associative', \emph{i.e.} $
\langle [X,Y],Z \rangle = \langle X,[Y,Z] \rangle,  \  \forall X,Y,Z \in \mathfrak{a}. 
$
Such a DD Lie algebra posseses a canonical quadratic Casimir
\begin{align}
C=\tfrac12\sum_{i}{(y^i\,Y_i+Y_i\,y^i)}.
\label{cascas}\nonumber
\end{align}
A given even-dimensional Lie group $G$ with Lie algebra $\mathfrak{a}$ can have several DD structures,  \emph{i.e.} several splittings of $\mathfrak{a}$ of the form~\eqref{agd} that are nonisomorphic, or no DD structures at all.

A Lie algebra $\mathfrak{g}$ endowed with a skew-symmetric cocommutator map
$
\delta:\mathfrak{g}\to \mathfrak{g}\otimes \mathfrak{g},
$
fulfilling the two following conditions:
\begin{itemize}
\item i) $\delta$ is a 1-cocycle, {\em  i.e.},
\be
\delta([X,Y])=[\delta(X),\,  Y\otimes 1+ 1\otimes Y] + 
[ X\otimes 1+1\otimes X,\, \delta(Y)] ,\qquad \forall \,X,Y\in
\mathfrak{g}.
\label{1cocycle}
\ee
\item ii) The dual map $\delta^\ast:\mathfrak{g}^\ast\otimes \mathfrak{g}^\ast \to
\mathfrak{g}^\ast$ is a Lie bracket on $\mathfrak{g}^\ast$,
\end{itemize}
is called a  {\em Lie bialgebra} ($\mathfrak{g},\delta$). There is a one-to-one correspondence between DD Lie algebras and Lie bialgebras on $\mathfrak{g}$, obtained by endowing the first Lie subalgebra $\mathfrak{g}$ with a skew-symmetric cocommutator map defined by the structure constants for $\mathfrak{g}^\ast$, namely
\be
[Y_i,Y_j]= c^k_{ij}Y_k, \qquad  \delta(Y_n)=f_{n}^{lm} Y_l\otimes Y_m.
\label{qqa}\nonumber
\ee
This correspondence is due to the fact that  the Jacobi identity for the DD Lie algebra~\eqref{agd} is equivalent to the 1-cocycle condition~\eqref{1cocycle} for the cocommutator map $\delta$ with respect to the Lie algebra $\mathfrak{g}$.

Moreover, each DD structure for a Lie algebra $\mathfrak{a}$ generates a solution $r$ of the classical Yang-Baxter equation (CYBE) on $\mathfrak{a}$, which is provided by the canonical  classical
$r$-matrix
\be
r=\sum_i{y^i\otimes Y_i} .
\label{rcanon}
\ee
Therefore, if $\mathfrak{a}$ is a finite-dimensional DD Lie algebra,
then it  can be endowed with the
(quasitriangular) coboundary Lie bialgebra structure $(\mathfrak{a},\delta_D)$ 
 given by
\be
\delta_D(X)=[ X \otimes 1+1\otimes X ,  r],
\quad
\forall X\in {\mathfrak{a}}.
\label{rcanon2}\nonumber
\ee
It is in terms of this canonical $r$-matrix that the Fock-Rosly condition (see~\cite{FR1999moduli,MS2003poisson,OS2018} and references therein) is found to be fulfilled~\cite{BHM2013cqg}.

To any coboundary Lie bialgebra, as the one defined by \eqref{rcanon}, a  {\em Poisson-Lie (PL)  group} structure $(G,\Pi)$ on $G$ is associated, and this PL structure is explicitly given by the so-called Sklyanin bracket 
\be
\{f,g\}= r^{ij} \left( \nabla^L_i f \nabla^L_j g - \nabla^R_i f \nabla^R_j g \right), \qquad f,g \in \mathcal C ^\infty(G),
\label{sklyanin}
\ee
where $r^{ij}$ are the elements of the skew-symmetric part of \eqref{rcanon} and $\nabla^L_i,\nabla^R_i$ the left- and right-invariant vector fields on the Lie group $G$. Moreover, for $G$ connected and simply connected, this structure is unique \cite{Drinfeld1983hamiltonian}. In the same manner as we can construct smooth manifolds from quotients of Lie groups by certain subgroups, a similar construction exists for PL groups where now the group action must be compatible with the Poisson structure. 

A  {\em Poisson homogeneous space} (PHS) of a PL group $(G,\Pi)$ is a Poisson manifold $(M,\pi)$ endowed with a transitive group action $\rhd: G\times M\to M$ which is a Poisson map with respect to the Poisson structure $\pi$ on $M$ and the product of the Poisson structures $\Pi \times \pi$ on $G \times M$. In particular, we are interested in coset spaces $M = G/H$ of the group $G$ with isotropy subgroup $H$, whose Lie algebra will be denoted by $\mathfrak{h}$. In the rest of the paper we will describe PL structures on $G$ that can be canonically projected to the coset $G/H$, and this is guaranteed if the so-called  {\em coisotropy condition} for the isotropy subalgebra $\mathfrak h$ is satisfied, namely
\be
\delta(\mathfrak h) \subset \mathfrak h \wedge \mathfrak g.
\label{coisotropy}
\ee
Furthermore, if $H$ is a Poisson subgroup with respect to $\Pi$ then $\mathfrak{h}$ is a sub-Lie bialgebra structure and it satisfies $\delta\left(\mathfrak{h}\right) \subset \mathfrak{h} \wedge \mathfrak{h}$. This is a stronger condition and for some purposes the resulting spaces can have better behaviour, however the standard coisotropy condition (\ref{coisotropy}) will be enough in order to have a well-defined PHS that can be obtained by projecting the Sklyanin bracket~(\ref{sklyanin}).

\section{Drinfel'd double structures for the Poincar\'e group}

In this section we review the relevant outcomes of the recent work \cite{BGH2018poincareDD}, in which the Poincar\'e case was thoroughly analyzed. From now on, we will work on a kinematical basis   $\{J, K_1, K_2, P_0, P_1, P_2\}$ corresponding  to the generators of rotations, boosts, time translation and space translations, respectively. This basis is specially adapted to describe the role of the Poincar\'e algebra as the algebra of isometries of Minkowski spacetime. In this basis the commutation relations read
\begin{align} 
\begin{array}{lll}
 [J,K_1]=K_2, & \quad [J,K_2]=-K_1,  & \quad  [K_1, K_2]=-J,  \\[2pt]
 [J,P_0]=0 ,& \quad [J,P_1]=P_2 ,& \quad [J, P_2]=-P_1 ,\\[2pt]
[K_1,P_0]=P_1 ,& \quad [K_1,P_1]=P_0 ,& \quad [K_1, P_2]=0,\\[2pt]
[K_2,P_0]=P_2 ,& \quad[K_2,P_1]= 0 ,& \quad[K_2, P_2]=P_0,\\[2pt]
[P_0,P_1]=0, & \quad[P_0, P_2]=0 ,& \quad  [P_1,P_2]=0 .
 \end{array}
\label{aaxx}\nonumber
\end{align}
This Lie algebra has two quadratic Casimir elements, namely
\be
C_1=P_0^2 - P_1^2 - P_2^2,\qquad
C_2=  J\,P_0 +K_2\,P_1 -  K_1\,P_2   .
\label{xx}\nonumber
\ee
In this notation, the (2+1)-dimensional Minkowski spacetime $M^{2+1}$ arises as the homogeneous space of the Poincar\'e group $ISO (2,1)\equiv P(2+1)$ having the Lorentz subgroup $H=SO (2,1)$ as the isotropy subgroup of the origin, that is,  $M^{2+1}\equiv  ISO (2,1)/SO (2,1)$. Therefore we have that  $
\mathfrak{a}=\mathfrak{iso} (2,1 )=\mathfrak{p}(2+1)=\spn\{ J,K_1,K_2,P_0,P_1,P_2 \} 
$ and
$ \mathfrak{h}={\rm Lie}(H)=\mathfrak{so} (2,1 )=\spn\{ J,K_1,K_2 \}.
$
In the classical case, the above coset construction induces a unique smooth structure on the coset manifold. However, the situation is quite different if we allow quantum deformations of the isometry groups because these deformations are far from being unique and so the resulting quantum (noncommutative) spacetimes are different. Of course as smooth manifolds they are diffeomorphic, but the induced Poisson structures are not equivalent, and will provide the semiclassical limit of the noncommutative spacetimes. Some of these Poisson Minkowski spacetimes have the extra property of being defined by an $r$-matrix that comes from a DD structure of the Poincar\'e Lie group and they are specially interesting due to their connections to (2+1) quantum gravity, as explained above. Here we sketch the main results given in \cite{BGH2018poincareDD}, where a complete study of this matter was presented. 

We recall that in \cite{Stachura1998} all the PL structures on the Poincar\'e Lie group were classified in terms of its defining $r$-matrix (given that for this group every PL structure is a coboundary one), and it was found that there exist eight (possibly multiparametric) classes of PL structures. On the other hand, there also exist eight non-isomorphic DDs on the Poincar\'e Lie group, each of them giving a canonical $r$-matrix and thus a different PL structure. In Table \ref{table:2+1_r-matrices} these DD $r$-matrices are presented together with the Poisson subgroup/coisotropy condition for each of them and the class of the classification \cite{Stachura1998} they belong to.


\begin{table}[tp]{\footnotesize
\caption{\small  The (2+1) Poincar\'e $r$-matrices and Poisson subgroup/coisotropy condition for the eight DD structures on $\mathfrak{p}(2+1)$ \cite{BGH2018poincareDD} as well as the corresponding class in the Stachura classification.} 
\label{table:2+1_r-matrices}
\begin{center}
\begin{tabular}{cllc}
\hline 
 &  &   &       \\[-1.5ex]
Case &  Classical $r$-matrix  $r'_i$& $\delta_D \left(\mathfrak{h}\right)$ &Class~\cite{Stachura1998} \\[3pt]
\hline 
 &  &   &       \\[-1.5ex]
0 & $ \frac{1}{2} ( - P_0\wedge J - P_1 \wedge K_2 + P_2\wedge K_1)$ & $=0$& (IV) \\[5pt] 
1 & $K_1 \wedge J +K_1 \wedge K_2 +( - P_0 \wedge J - P_1 \wedge K_2 + P_2 \wedge K_1)$ & $\subset \mathfrak{h}\wedge \mathfrak{h}$ &  (I) \\[4pt] 
2 & $P_2 \wedge J - P_0 \wedge K_2 - P_2 \wedge K_2 + \frac12 (P_0 \wedge J - P_1 \wedge K_2 + P_2 \wedge K_1)$ & $\subset \mathfrak{h}\wedge \mathfrak{g}$ &  (IIa) \\[5pt]  
3 & $ - P_2 \wedge J - P_0 \wedge K_2 - P_2  \wedge K_2+ \frac{1}{2}(- P_0 \wedge J + P_1 \wedge K_2 - P_2 \wedge K_1)  $ & $\not\subset \mathfrak{h}\wedge \mathfrak{g}$   &(IIa) \\[4pt] 
  & $\qquad  + \frac{1}{\lambda}\bigl(P_0 \wedge P_1 + 2 (P_0 \wedge P_2 + P_2 \wedge P_1) \bigr)$ &    & \\[5pt] 
4 & {$P_2 \wedge J + \frac12 (P_0 \wedge J - P_1 \wedge K_2 + P_2 \wedge K_1) + \lambda P_0 \wedge P_2$} & {$\not\subset \mathfrak{h}\wedge \mathfrak{g}$} &(IIIb) \\[5pt]  
5 & $P_1 \wedge J + \frac{1}{2} \left( - P_0 \wedge J + P_1 \wedge K_2 - P_2 \wedge K_1 \right) + \frac{1}{\lambda} P_1 \wedge P_0$ & $\not\subset \mathfrak{h}\wedge \mathfrak{g}$ & (IIIb)\\[5pt]  
6 & $P_0 \wedge K_2 + \frac12 (- P_0 \wedge J + P_1 \wedge K_2 + P_2 \wedge K_1)$ & $\subset \mathfrak{h}\wedge \mathfrak{g}$ & (IIIb)\\[5pt]   
7 & $P_2 \wedge J + \frac12 (- P_0 \wedge J + P_1 \wedge K_2 - P_2 \wedge K_1)$ &  $\subset \mathfrak{h}\wedge \mathfrak{g}$ & (IIIb) \\[5pt] 
\hline
\end{tabular} 
\end{center}}
\end{table}


The first $r$-matrix in   Table~\ref{table:2+1_r-matrices} is the one corresponding to the `Lorentz double', i.e.~the one coming from the Lie bialgebra $(\mathfrak{g}=\mathfrak{so}(2,1)\simeq \mathfrak{sl}_2(\mathbb R), \delta\equiv 0)$ and the resulting Poisson Minkowski spacetime associated is of Lie algebraic type, in fact isomorphic to $\mathfrak{so}(2,1)$. Out of the other seven cases, four of them satisfy the coisotropy condition, although only the Case 1   is of Poisson subgroup type (as, obviously, the Lorentz double of Case 0). These five DDs fulfilling the coisotropy condition  give  rise to five different Poisson Minkowski spacetimes, which are displayed in Table \ref{table:2+1_Poisson_Minkowski-spacetimes}. We remark that Cases 1 and 2 are written as a two-parametric generalization of the corresponding DD structure, which is properly recovered for $\alpha_1=\beta_1$ and $\alpha_2=\beta_2$, respectively.


\begin{table}[tp]{\footnotesize
\caption{\small The (2+1) Poisson Minkowski spacetimes arising from coisotropic DD structures \cite{BGH2018poincareDD}.} 
\label{table:2+1_Poisson_Minkowski-spacetimes}
\begin{center}
\begin{tabular}{cccc}
\hline 
 &  &   &       \\[-1.5ex]
Case & $\{x^0,x^1\}$ & $\{x^0,x^2\}$ & $\{x^1,x^2\}$ \\[3pt]
\hline 
 &  &   &       \\[-1.5ex]
0 & $-x^2$ & $x^1$ & $x^0$  \\[5pt] 
1 & $-\alpha_1 x^2 (x^0 + x^1) +2 \beta_1 x^2$ & $\alpha_1 x^1 (x^0 + x^1) - 2 \beta_1 x^1$ & $\alpha_1 x^0 (x^0 + x^1) - 2 \beta_1 x^0$  \\[5pt] 
2 & $0$ & $-\alpha_2 (x^0-x^2)$ & $-\beta_2 (x^0-x^2)$  \\[5pt]  
6 & $0$ & $-x^0+x^1$ & $0$ \\[5pt]   
7 & $0$ & $0$ & $-(x^0+x^2)$ \\[5pt] 
\hline
\end{tabular} 
\end{center}}
\end{table}


Four of these five  Poisson Minkowski spacetimes are of Lie algebraic type (namely Cases 0, 2, 6 and 7) while the remaining Case 1 is the only quadratic one. Moreover, Cases 0 and 1 are the only ones that are of Poisson subgroup type, and Case 0 is the only one which is both of Lie algebraic and of Poisson subgroup type. These two conditions singularize this Poisson Minkowski spacetime from the rest and makes the construction of its associated  quantum Poincar\'e group an interesting problem. Looking at Table \ref{table:2+1_Poisson_Minkowski-spacetimes} we see that the Poisson Minkowski spacetime obtained from Case 1 is a quadratic generalization of Case 0. 

Further information on the structure of Poincar\'e DD $r$-matrices can be obtained by performing a contraction procedure on the $r$-matrices coming from the DD structures for the (anti-)de Sitter ((A)dS) groups, which were studied in \cite{BHM2013cqg}. In this way we obtain the answer to the question on which of the eight Poincar\'e DD $r$-matrices can be seen as the vanishing cosmological constant limit  of the (A)dS ones. The result of this analysis, performed in \cite{BGH2018poincareDD}, is that four of the seven different DD (A)dS $r$-matrices give rise to Poincar\'e DD $r$-matrices, namely Cases 0 and 2. Note that Case 0 is also special in this regard, while Case 2 is found to produce an $r$-matrix which is isomorphic to the `space-like' $\kappa$-Poincar\'e $r$-matrix along with a twist~\cite{BHM2014sigma}. In this respect, it should be noted that Table \ref{table:2+1_r-matrices} indicates that a similar DD construction of the twisted `time-like' $\kappa$-Poincar\'e deformation is not possible.

An interesting remark concerning DDs for the Poincar\'e group is that the (2+1)-dimensional situation is quite special, since in (3+1) and higher dimensions, the existence of any DD  is precluded by the non-existence of a suitable non-degenerate bilinear form \eqref{ages}, both symmetric and associative. In (1+1)-dimensions there is obviously no DDs (given that the group has odd dimension), but it turns out that the non-trivially centrally extended Poincar\'e group in (1+1) dimensions has two non-isomorphic DD structures which were also explicitly constructed in \cite{BGH2018poincareDD}.


\section{Drinfel'd double Euclidean $r$-matrices and Poisson homogeneous spaces}

Once the complete set of DDs for the Poincar\'e group has been clarified, it certainly makes sense to study the DDs for the Euclidean group in 3-dimensions, both because of its inherent interest related with the construction of Poisson Euclidean spaces and also for the possibility of comparing DD structures  for these two closely related Lie algebras. So let us consider  the  Euclidean  Lie algebra $\mathfrak{e}(3)=\mathfrak{iso}(3)$ in terms of generators of rotations $J_i$ and translations $P_i$ ($i=1,2,3)$.  The commutation rules  read
 \begin{align} 
\begin{array}{l}
[J_i,J_j]=\varepsilon_{ijk} J_k,\qquad [J_i,P_j]=\varepsilon_{ijk} P_k,\qquad [P_i,P_j]=0,\qquad i,j,k=1,2,3,
 \end{array}
\label{aa}
\end{align}
where $\varepsilon_{ijk} $ is the totally skewsymmetric tensor with $\varepsilon_{123} =1$.  The two quadratic Casimir elements for this algebra are given by
\be
C_1=P_1^2 +P_2^2 + P_3^2,\qquad
C_2= J_1\,P_1 +J_2\,P_2+ J_3\,P_3 .
\label{pcasimirs}
\ee

The Euclidean   space in three dimensions, $E^{3}$, can be constructed   as the homogeneous space of the Euclidean isometry group $ISO (3 )=E(3)$ having the   subgroup $H=SO (3 )$ as the isotropy subgroup of the origin, that is,   $E^{3}\equiv  ISO (3 )/SO (3 )$. Hence we have that
$
\mathfrak{a}= \mathfrak{iso} (3 )=\mathfrak{e}(3)=\spn\{ J_1,J_2,J_3,P_1,P_2,P_3 \},$
 and
 $ \mathfrak{h}={\rm Lie}(H)=\mathfrak{so} (3 )=\spn\{J_1,J_2,J_3 \}.
$
If we denote $\mathfrak{t} =\spn\{P_1,P_2,P_3 \}$, we have that $\mathfrak{e}(3) = \mathfrak{h} \oplus \mathfrak{t}$ as a vector space.

According to the classification \cite{Gomez2000} there is no `non-trivial' DD structure for $E(3)$. However, $E(3)$ has the `trivial' DD structure induced by its semidirect product form. Notice that $E(3) = SO(3) \ltimes \mathbb{R}^3$ is the semidirect product of the rotation subgroup and the translations, inherited by its Lie algebra $\mathfrak{e}(3) = \mathfrak{so}(3) \oplus_S \mathbb{R}^3$, and this is just the DD structure arising in correspondence with the Lie bialgebra structure $(\mathfrak{g},\delta) = (\mathfrak{so}(3),\delta \equiv 0)$. It is straightforward to check that such unique DD structure for $\mathfrak{e}(3)$ is given by the isomorphism 
$Y_i = J_i$ and $ y^i = P_i $, 
and thus we have $c_{ij}^k = \varepsilon_{ijk}$ and $f^{ik}_j=0$. In this way we obtain
\begin{equation}
r=\sum_{i }  P_i \otimes J_i \, ,   \qquad C_2 = \sum_{i }  J_i \otimes P_i \, ,
\nonumber
\end{equation}
so the skew-symmetric component of the $r$-matrix reads
\begin{equation}
r' = r - C_2 = \sum_{i }  P_i \wedge J_i,
\label{dde}
\end{equation}
while the induced pairing has as non-vanishing entries
$
\langle P_i,J_i \rangle = 1.
$
This DD structure is the Euclidean analogue of the Case 0 in the Poincar\'e clasification, and is directly related to the semidirect product structure of both Lie groups. The striking difference between the Euclidean and the Poincar\'e Lie groups is that while in the latter there is a plurality of DD structures (eight non-isomorphic ones), in the former a single one   does exist.

\section{Classification of the coboundary Poisson Euclidean spaces}

In~\cite{Stachura1998} the complete classification of $r$-matrices for the three-dimensional Euclidean Lie algebra was obtained. 
In order to relate such a classification with our results we rename the Euclidean generators in~\cite{Stachura1998} as
$ e_i = P_i$ and   $ k_i = J_i $ $( i=1,2,3)$ so
satisfying the commutation rules (\ref{aa}). In such classification, the Euclidean classical $r$-matrices are expressed in the form
$
r=a+b+c,
$
with
\be
a\subset \mathfrak t \wedge \mathfrak t,
\qquad
b\subset \mathfrak t \wedge \mathfrak h,
\qquad
c\subset \mathfrak h \wedge \mathfrak h,
\nonumber
\ee
where $\mathfrak t =\spn\{P_1,P_2,P_3 \}$ and $\mathfrak h =\spn\{ J_1,J_2,J_3\}= \mathfrak{so}(3)$.
It turns out that the classification is ruled by two real parameters $\mu$ and $p$ according to the following relations:
\be
[[a,b]] = p\, \tilde \eta, \quad 2 [[a,c]] + [[b,b]] = \mu \, \Omega, \quad [[b,c]]=[[c,c]]
=0,\quad  \tilde \eta \in \bigwedge^3 \mathfrak{t}, \quad \Omega \in \bigwedge^2 \mathfrak{t} \otimes \mathfrak{h},
\nonumber
\ee
where $ [[\cdot \, ,\cdot]] $ denotes the Schouten bracket. 
The cases with $p=0$ correspond to $r$-matrices leading to coisotropic Lie bialgebras with respect to $ \mathfrak{h} =\spn\{J_1,J_2,J_3 \}$.  
It turns out that there exist three equivalence classes of  three-dimensional Euclidean $r$-matrices, all of them with $c=0$, which in our notation read: 
 
 \medskip
 
\noindent{\bf Class (I)}
\be
b= \alpha (P_1 \wedge J_2 - P_2 \wedge J_1)+ \rho P_3 \wedge J_3,
\qquad
a=a_{12} P_1\wedge P_2 + a_{13} P_1\wedge P_3 + a_{23} P_2\wedge P_3 ,
\label{classIIa}\nonumber
\ee
with $\alpha=\{0,1\}$, $\rho \ge 0$, $\alpha^2+\rho^2\ne 0$, $\mu=-2\alpha^2$, $p\in\mathbb{R}$,  and where from now on  $\{a_{12},a_{13},a_{23}\}$ denote  three  free real parameters. 

\medskip

\noindent{\bf Class (II)}
\be
b=  P_1 \wedge J_1+  P_2\wedge J_2 +P_3\wedge J_3  , \qquad a=0,
\label{classII}
\ee
with  $\mu=2$ and $p=0$. 

 \medskip
 
 \noindent{\bf Class (III)}
\be
b=0,
\qquad a=a_{12} P_1\wedge P_2 + a_{13} P_1\wedge P_3 + a_{23} P_2\wedge P_3 ,
\label{classIII}
\ee
with  $\mu=0$ and $p=0$. 

\medskip

With this classification at hand, we can easily  identify the only three-dimensional Euclidean DD $r$-matrix (\ref{dde})  with the one in Class (II) (\ref{classII})  in~\cite{Stachura1998}. Quite interestingly, exactly as it happened for the Poincar\'e case, the `trivial' DD $r$-matrix is the one corresponding to the only non-parametric family of coboundary PL structures on the three-dimensional Euclidean group. 

Regarding different dimensions, no DD structure  exists for the Euclidean group. In higher dimensions than three, this is due to the lack of existence of a non-degenerate associative symmetric bilinear form. In the two-dimensional case, the statement follows because there are only three non-isomorphic DDs \cite{Gomez2000}, two of them isomorphic to the non-trivially centrally extended Poincar\'e group and the other one isomorphic to the non-trivially centrally extended AdS group. Therefore, no centrally extended Euclidean group can be endowed with a DD.

\section{Contraction of  Drinfel'd double $r$-matrices from $\mathfrak{so}(3,1)$}

The complete study of DDs for the Lie algebra $\mathfrak{so}(3,1)$ was carried out   in~\cite{BHM2013cqg}.
In what follows we analyze the contraction of such structures to the Euclidean case. This contraction procedure can be understood in geometric terms as the zero-curvature limit of the three-dimensional  hyperbolic space whose isometry group is just ${SO}(3,1)$.
 
The classification of  $\mathfrak{so}(3,1)$ DD $r$-matrices in~\cite{BHM2013cqg} was performed in the usual Chern-Simons basis $\{J_0,J_1,J_2,P_0,P_1,P_2\}$. 
It turns out that there are four DDs for $\mathfrak{so}(3,1)$, called cases A, B, C and D in~\cite{BHM2013cqg}.  
For the cases  A and C     the relationship with the geometrical basis used throughout the present paper is established by means of the     isomorphism  given by
\be
J_0 \rightarrow -J_1, \quad J_1 \rightarrow \tfrac{1}{\kk}\, P_3,\quad J_2 \rightarrow \tfrac{1}{\kk}\, P_2  , \quad P_0 \rightarrow P_1, \quad P_1 \rightarrow \kk \,J_3, \quad  P_2 \rightarrow \kk \,J_2, 
\label{csiso}
\ee
where $\kk$ is a non-zero real parameter. And for the cases B and D the  isomorphism reads
\be
 J_s \rightarrow J_{s+1},  \qquad  P_s\rightarrow P_{s+1}, \qquad  s=0,1,2.
\label{csiso2}
\ee
By applying these two isomorphisms, we  find that the commutation relations for the  Lie algebra $\mathfrak{so}(3,1)$ adopt the form
\begin{align} 
\begin{array}{l}
[J_i,J_j]=\varepsilon_{ijk} J_k,\qquad [J_i,P_j]=\varepsilon_{ijk} P_k,\qquad [P_i,P_j]=-\kk^2 \varepsilon_{ijk} J_k,\qquad i,j,k=1,2,3.
 \end{array}
\label{bb}
\end{align}
In this basis, the two quadratic Casimir elements for $\mathfrak{so}(3,1)$ can be written as
\be
C_1=P_1^2 +P_2^2 + P_3^2-\kk^2\bigl( J_1^2+J_2^2+J_3^2\bigr),\qquad
C_2= J_1\,P_1 +J_2\,P_2+ J_3\,P_3 .
\label{bc}
\ee
Now we consider the three-dimensional hyperbolic space as the  homogeneous space of the   isometry group $SO (3 ,1)$ with  isotropy  subgroup $H=SO (3 )$,   $H^{3}\equiv  SO (3,1 )/SO (3 )$, provided that
$ 
\mathfrak {a}= \mathfrak{so} (3,1 )= \spn\{ J_1,J_2,J_3,P_1,P_2,P_3 \}$ and $ \mathfrak{h}={\rm Lie}(H)=\mathfrak{so} (3 )=\spn\{J_1,J_2,J_3 \}.
$
The  hyperbolic space $H^{3}$ has negative constant sectional curvature equal to $-\kk^2$, so that the parameter $\kk$ is related with the radius of the space $R$ through $\kk=1/R$. The `flat' contraction to the Euclidean algebra and space thus corresponds to applying the limit $\kk\to 0$ ($R\to \infty$). In this manner, the commutation rules (\ref{bb}) and Casimirs (\ref{bc}) 
reduce to the Euclidean ones (\ref{aa}) and (\ref{pcasimirs}), respectively.

 Next,  by using (\ref{csiso}) and (\ref{csiso2}) in the results given in~\cite{BHM2013cqg}  we obtain   the following four DD  $r$-matrices  for $\mathfrak{so}(3,1)$:
\begin{align} 
\begin{array}{l}
\displaystyle{  r'_{\rm A}= \tfrac 1\kk \, P_3\wedge P_2+\tfrac 12 (P_1\wedge J_1 - P_2\wedge J_2- P_3\wedge J_3),  }\\[3pt]
\displaystyle{     r'_{\rm B}=-\kk J_2\wedge J_3+\tfrac 12  (P_1\wedge J_1 + P_2 \wedge J_2 + P_3 \wedge J_3 )
     , }\\[3pt]
  \displaystyle{      r'_{\rm C}= \tfrac 12 \left( \tfrac 1\kk \, P_3\wedge P_1+\kk J_1\wedge J_3 +P_2\wedge J_2 \right),}  \\[3pt]
\displaystyle{          r'_{\rm D}=J_1\wedge P_2 - J_2 \wedge P_1 +\frac{(1+\mu^2)}{2\mu} \, P_3\wedge J_3 + \frac{(\mu^2-1)}{2\kk\mu}\left( \kk^2 J_1\wedge J_2 - P_1\wedge P_2\right) ,\quad \mu>0.} \end{array}
\label{bcc}
\end{align}
In principle, only the $r$-matrix $ r'_{\rm B}$ has a well defined flat limit $\kk\to 0$. Nevertheless, we can scale the remaining cases in order to obtain four contracted Euclidean $r$-matrices; these are
\bea
&&  \lim_{\kk\to 0}\kk \, r_{\rm A}' =P_3\wedge P_2,\qquad  \lim_{\kk\to 0}  r_{\rm B}' =\tfrac 12 (P_1\wedge J_1 + P_2 \wedge J_2 + P_3 \wedge J_3) , \cr
&&   \lim_{\kk\to 0}\kk \, r_{\rm C}' =\frac 12 \,P_3\wedge P_1,\qquad    \lim_{\kk\to 0}\kk \, r_{\rm D}' = \frac{(1-\mu^2)}{2 \mu} \, P_1\wedge P_2 ,\quad \mu>0 .
\label{bd}\nonumber
\eea
Consequently, the cases A, C and D give rise to Euclidean $r$-matrices belonging to Class (III) (\ref{classIII}) of~\cite{Stachura1998}, meanwhile the case B gives exactly the complete Class (II) (\ref{classII}). Note that this Class is the one obtained from the DD structure for the Euclidean group (\ref{dde}). We recall that in \cite{BGH2018poincareDD},  it was found that four (A)dS (two for dS and two for AdS) DD $r$-matrices contract to two DD Poincar\'e $r$-matrices (Cases 0 and 2).

We remark that the initial pairing for the cases A, B and C for the $\mathfrak{so}(3,1)$ DDs in the basis (\ref{bcc}) has the non-vanishing entries
$\langle P_i,J_i \rangle = 1$, which does not depend on $\kk$, so that it remains unchanged under contraction. In contrast, the pairing for case D diverges under the limit $\kk\to 0$.


\section{Poisson homogeneous Euclidean spaces}

In the previous sections we have studied the   DD structures  for the Euclidean group in three dimensions and we have identified to which of the coboundary PL structures they correspond. Now we present the full construction of Poisson homogeneous Euclidean spaces, based in the classification presented in section 5. In order to perfom that, we first need to introduce suitable coordinates on the Euclidean   group and we write a generic element $Q$ of the Lie algebra $\mathfrak{e}(3)$ as
\begin{equation}
Q = x^1 P_1 + x^2 P_2 + x^3 P_3 + \theta^1 J_1 + \theta^2 J_2 + \theta^3 J_3= 
\begin{pmatrix}
0 & 0 & 0 & 0 \\
x^1 & 0 & -\theta^3 & \theta^2 \\
x^2 & \theta^3 & 0 & -\theta^1 \\
x^3 & -\theta^2 & \theta^1 & 0
\end{pmatrix}.
\nonumber
\end{equation}
Next we  introduce the coordinates on the Lie group as the ones associated to each Lie algebra generator through the exponential map
\begin{equation}
g=\exp(x^1 P_1) \exp(x^2 P_2) \exp(x^3 P_3) \exp(\theta^1 J_1) \exp(\theta^2 J_2) \exp(\theta^3 J_3),
\nonumber
\end{equation}
and then we compute left- and right-invariant vector fields in these coordinates, thus obtaining 
{\small
\bea
&& \!\!\!\!\!\!\!\!\!\!\!\!\!\! X^L_{J_1} = \frac{\cos \theta^3}{\cos \theta^2} \left( \partial_{\theta^1}- \sin \theta^2 \partial_{\theta^3} \right) + \sin \theta^3 \partial_{\theta^2}, \quad
X^L_{J_2} = \frac{\sin \theta^3}{\cos \theta^2} \left( -\partial_{\theta^1} + \sin \theta^2 \partial_{\theta^3} \right) + \cos \theta^3 \partial_{\theta^2},  \quad
  X^L_{J_3} = \partial_{\theta^3}, \nonumber\\
&& \!\!\!\!\!\!\!\!\!\!\!\!\!\! X^L_{P_1} = \cos \theta^2 \cos \theta^3 \partial_{x^1} + \left( \sin \theta^1 \sin \theta^2 \cos \theta^3 + \cos \theta^1 \sin \theta^3 \right) \partial_{x^2} - \left(  \cos \theta^1 \sin \theta^2 \cos \theta^3 - \sin \theta^1 \sin \theta^3 \right) \partial_{x^3},  \nonumber\\
&& \!\!\!\!\!\!\!\!\!\!\!\!\!\!  X^L_{P_2} = -\cos \theta^2 \sin \theta^3 \partial_{x^1} - \left( \sin \theta^1 \sin \theta^2 \sin \theta^3 - \cos \theta^1 \cos \theta^3 \right) \partial_{x^2} + \left( \cos \theta^1 \sin \theta^2 \sin \theta^3 + \sin \theta^1 \cos \theta^3 \right) \partial_{x^3},  \nonumber\\
&& \!\!\!\!\!\!\!\!\!\!\!\!\!\! X^L_{P_3} = \sin \theta^2 \partial_{x^1} - \cos \theta^2 \left(  \sin \theta^1 \partial_{x^2} - \cos \theta^1 \partial_{x^3} \right), 
\nonumber
\eea
}
{\small
\begin{equation}
\begin{alignedat}{2}
& X^R_{J_1} = -x^3 \partial_{x^2} + x^2 \partial_{x^3} + \partial_{\theta^1}, \qquad & X^R_{P_1} = \partial_{x^1}, \\
& X^R_{J_2} = x^3 \partial_{x^1} - x^1 \partial_{x^3} + \cos \theta^1 \partial_{\theta^2} + \frac{\sin \theta^1}{\cos \theta^2} \left( \sin \theta^2 \partial_{\theta^1} - \partial_{\theta^3} \right), \qquad & X^R_{P_2} = \partial_{x^2}, \\
& X^R_{J_3} = -x^2 \partial_{x^1} + x^1 \partial_{x^2} + \sin \theta^1 \partial_{\theta^2} + \frac{\cos \theta^1}{\cos \theta^2} \left( -\sin \theta^2 \partial_{\theta^1} + \partial_{\theta^3} \right), \qquad & X^R_{P_3} = \partial_{x^3}.
\end{alignedat}
\nonumber
\end{equation}}
\noindent
Thus we have all the ingredients to  study explicitly the Poisson homogeneous Eucliean spaces.  The final result is as follows.
\smallskip

\noindent{\textbf{Class (II).}}
This is the only $r$-matrix coming from a DD. Its cocommutator reads
\begin{equation}
\delta (J_i) = 0, \qquad
\delta (P_1) = 2 P_2 \wedge P_3, \qquad
\delta (P_2) = -2 P_1 \wedge P_3, \qquad
\delta (P_3) = 2 P_1 \wedge P_2, 
\nonumber
\end{equation}
which shows that its associated PHS is of Poisson subgroup type in a trivial way (this is consistent with the fact that the  $r$-matrix (\ref{classII})  is  the analogue of the Poincar\'e Case 0 studied in \cite{BGH2018poincareDD}, see Table~\ref{table:2+1_r-matrices}). The associated PHS is given by the fundamental Poisson bracket
\begin{equation}
\{ x^i, x^j \} = 2 \varepsilon_{ijk} x^k,
\label{ncdd}
\end{equation}
which is  so isomorphic to the $\mathfrak{so}(3)$ Lie algebra, while its Poincar\'e counterpart was isomorphic to $\mathfrak{so}(2,1)$, as shown in Table \ref{table:2+1_Poisson_Minkowski-spacetimes}. As a matter of fact, if we compute the full Sklyanin bracket we get that the remaining   group coordinates Poisson commute $
\{ x^i, \theta^j \} = \{ \theta^i, \theta^j \} = 0.
$

\smallskip
 
\noindent{\textbf{Class (III).}}
This family of $r$-matrices are solutions of the CYBE, and its cocommutator reads
\begin{equation}
\begin{split}
\delta_a (J_1) &= -a_{13} P_1 \wedge P_2 + a_{12} P_1 \wedge P_3, \\
\delta_a (J_2) &= -a_{23} P_1 \wedge P_2 + a_{12} P_2 \wedge P_3, \qquad \delta_a (P_i) = 0, \\
\delta_a (J_3) &= -a_{23} P_1 \wedge P_3 + a_{13} P_2 \wedge P_3. 
\end{split}
\nonumber
\end{equation}
This cocommutator is not coisotropic with respect to the isotropy subgroup of rotations (apart from the trivial case $r=a=0$), and we shall not write down the Poisson brackets for the group coordinates as we are only interested in describing coisotropic PHS. 

\smallskip
 
\noindent{\textbf{Class (I).}}
This multiparametric family of $r$-matrices is composed by solutions of the form of Class (III) plus some new terms. They satisfy the CYBE iff $\alpha=0$. The cocommutator reads
\begin{equation}
\begin{split}
\delta (J_1) &= \alpha (-P_3 \wedge J_1 + P_1 \wedge J_3) - \rho (P_3 \wedge J_2 + P_2 \wedge J_3) + \delta_a (J_1), \\
\delta (J_2) &= \alpha (-P_3 \wedge J_2 + P_2 \wedge J_3) + \rho (P_3 \wedge J_1 + P_1 \wedge J_3) + \delta_a (J_2), \qquad
\delta (J_3) = \delta_a (J_3), \\
\delta (P_1) &= \alpha  P_1 \wedge P_3 + \rho P_2 \wedge P_3, \qquad
\delta (P_2)  = \alpha P_2 \wedge P_3 - \rho P_1 \wedge P_3, \qquad
\delta (P_3) = 0, 
\end{split}
\nonumber
\end{equation}
proving that they are coisotropic deformations if the $a$-terms vanish, but they are never  of Poisson subgroup type. So, in the case with $a=0$ the associated Poisson Euclidean spaces read
\begin{equation}
\begin{split}
\{ x^1, x^2 \} = 0,  \qquad
\{ x^1, x^3 \} = \alpha x^1 - \rho x^2, \qquad
\{ x^2, x^3 \} = \alpha x^2 + \rho x^1. 
\end{split}
\nonumber
\end{equation}
The distinguished behavior of the third coordinate becomes evident both in the cocommutator and in the Poisson bracket, in contrast what happens in the DD noncommutative space~(\ref{ncdd}).


\section{Concluding remarks}

Since the complete construction of DDs for the Poincar\'e Lie group was recently presented in \cite{BGH2018poincareDD}, it seemed  natural to perform the analogous analysis for the three-dimensional Euclidean group, since all these DDs are relevant for (2+1) quantum gravity, both in its Lorentzian and Euclidean facets. Also, in this paper we have presented the construction of the full set of Poisson homogeneous Euclidean spaces, and it has  been  shown how two of the three classes of coboundary PL structures on the Euclidean group generate coisotropic Poisson homogeneous spaces. 

Regarding DDs, it is worth stressing that the large plurality of nonisomorphic DDs for the Poincar\'e group is lost in its Euclidean counterpart, and this is clearly due to the flexibility of the  Lorentz sector in order to give rise to DDs. However, while differences are striking it should be noted that the most studied DD structure (the `Lorentz double'~\cite{MW1998}) has an Euclidean analogue (the `rotation double' or the `$\mathfrak{su}(2)$ double'~\cite{BM1998topological, BM2003, Majid2005time, JMN2009}) with similar algebraic properties, and both   are  canonically induced by the semidirect product structure of the group of isometries. Also, it should be noted that  the corresponding contraction procedures applied to the (A)dS  $ r$-matrices gives similar results for the Lorentz and rotation doubles.


\section*{Acknowledgments}

This work was partially supported by Ministerio de Ciencia, Innovaci\'on y Universidades (Spain) under grant MTM2016-79639-P (AEI/FEDER, UE), by Junta de Castilla y Le\'on (Spain) under grant BU229P18 and by the Action MP1405 QSPACE from the European Cooperation in Science and Technology (COST). IG-S acknowledges a predoctoral grant from Junta de Castilla y Le\'on and the European Social Fund.



\end{document}